\documentstyle[aps,twocolumn]{revtex}

\begin{document}
\draft
\title{Strong eigenfunction correlations near the Anderson
localization transition.}

\author{Yan V. Fyodorov$^{a,c}$ and Alexander D. Mirlin $^{b,c}$}
\address{$^a$Fachbereich Physik, Universit\"{a}t-GH Essen, Essen
45117,Germany}
\address{$^b$Institut f\"ur Theorie der Kondensierten Materie,
         Universit\"at
        Karlsruhe, 76128 Karlsruhe,Germany}
\address{$^c$ Petersburg
Nuclear Physics Institute, Gatchina 188350, Russia}

\date{ December 24, 1996}
\maketitle

\begin{abstract}
We study overlap of two different eigenfunctions as compared with
self-overlap in the framework of an infinite-dimensional version of
the disordered tight-binding model. Despite a very sparse structure of
the eigenstates in the vicinity of Anderson transition
their mutual overlap is still found to be of the same order as
self-overlap as long as energy separation is smaller than a critical
value.
The latter fact explains robustness of the Wigner--Dyson level
statistics everywhere in the phase of extended states.
The same  picture is expected
to hold for usual $d-$dimensional conductors, ensuring
the  $s^{\beta}$  form  of the  level repulsion  at  critical point.
\end{abstract}

Recently, there was considerable growth of interest towards
constructing a unified picture of wave functions and energy levels
statistics for disordered conductors in the vicinity of Anderson
metal-insulator transition
\cite{AS2,Chalk,Shk,Huck,Eva,Braun,Zhar,Krav,AM,CLS,CKL,Pracz}.
Coming from the metallic phase, a typical wavefunction $\Psi_i({\bf
r})$ is extended and covers all the sample volume
randomly, but uniformly. When system approaches the point of
Anderson transition $E_c$, these extended eigenfunctions become less and
less homogeneous in space showing regions with larger and smaller
amplitudes and eventually forming a multifractal structure in the
vicinity of $E_c$.

To characterize the degree of
non-homogeneity quantitatively it is convenient to use the inverse
participation ratio (IPR)
$ I(E)=\int d{\bf r}\alpha({\bf r},E)$, where
\begin{equation}
 \alpha({\bf r},E)=\langle|\Psi_i({\bf r})|^4\rangle_E\equiv
\Delta\left\langle\sum_i|\Psi_i({\bf r})|^4\delta(E-E_i)\right\rangle\ ,
\label{1a}
\end{equation}
$\Delta$ is the mean level spacing
and the  angular brackets stand for the disorder averaging.
For extended states this quantity is
inversely proportional to the system volume: $I(E)=C(E)
L^{-d}$, with $L$ and $d$ standing for the system size and spatial
dimension, respectively. The coefficient $C$ in this relation
measures a fraction of the system volume where eigenfunction is
appreciably non-zero. For random homogeneous states $C\sim 1$,
whereas close to the mobility edge $E=E_c$ it becomes large
and diverges like
$C(E)\propto |E-E_c|^{-\mu},\quad \mu>0$ \cite{Weg} signalizing of  
increasing
sparsity of eigenfunctions. Just at the mobility edge eigenfunctions
occupy a vanishing fraction of the system volume and IPR scales like
$I(E)\propto L^{-d+\eta}, \eta>0$. Such a behavior reflects
fractal (actually, {\it multifractal}\cite{CdC,Chalk,Huck}) structure
of critical
eigenstates. At last, in the insulating phase any eigenstate
is concentrated in a domain of finite extension $\xi_l$ and IPR stays
finite in the limit of infinite system size $L\to \infty$.

This transparent picture serves as a basis for qualitative
understanding of spectral properties of disordered conductors.
Indeed, as long as eigenstates are well extended they overlap
substantially and corresponding energy levels repel each other
in
the same way as do eigenvalues of large random matrices studied by
Wigner and Dyson. As a result, the Wigner-Dyson (WD) statistics describes
well energy levels in a good metal\cite{Efrev,AShk}.
In contrast, in the insulating phase different eigenfunctions
corresponding to levels close in energy are localized far apart from
one another and their overlap is negligible. This is the reason for
absence of correlations of energy levels in this regime -- the
so-called Poisson statistics.

However, close to the transition point such a reasoning should be
used with caution. Naively one may expect that sparse (multifractal  
in the
critical point) eigenstates fail to overlap, that would result in
essential
weakening of level correlations close to the mobility edge and
vanishing level repulsion at $E=E_c$. However, a thorough
investigation shows \cite{AS2,Shk,Eva,Braun,Zhar,Krav,AM} that even
at the mobility edge levels
repel each other strongly, though the whole statistics is
 different from the WD one. One of the main purposes of the
present paper is to resolve this apparent contradiction.
We will  show that critical eigenstates for
nearby levels are strongly correlated and overlap well in spite of
their sparse structure.

The overlap of two different eigenstates $\Psi_i$ and $\Psi_j$
 corresponding to energy separation $|E_i-E_j|=\omega$
can be  characterized by comparing the correlation function
\begin{eqnarray}
&&\sigma(\bbox{r},E,\omega)=\left\langle|\Psi_i({\bf r})|^2
  |\Psi_j({\bf r})|^2\right\rangle|_{E,\omega}
\equiv \Delta^2
R_2^{-1}(\omega)\nonumber \\
&&\times \left\langle\sum_{i,j}|\Psi_i({\bf r})|^2
|\Psi_j({\bf r})|^2
\delta(E-E_i)\delta(E+\omega-E_j)\right\rangle
\label{1}
\end{eqnarray}
at $\omega\ne 0$ with  $\alpha(\bbox{r},E)$, the latter function playing 
in such a context the role of the eigenfunction self-overlap.
Here $R_2(\omega)$ denotes the two-level correlation function,
\begin{equation}\label{3}
R_2(\omega)=\Delta^2\left\langle\sum_{ij}\delta(E-E_i)\delta(E+\omega-E_j)
\right\rangle\end{equation}
To study the function $\sigma(\bbox{r},E,\omega)$
analytically, we consider an exactly solvable model of
the Anderson transition -- so-called sparse random matrix (SRM)
model \cite{sp1,sp2}. This model deals with large $N\times N$ matrices
(real symmetric or Hermitian) whose entries $H_{ij}$ are independent
random variables characterized by the following probability
distribution:
\begin{equation}\label{2}
{\cal P}(H_{ij})=(1-p/N)\delta(H_{ij})+\left(p/N\right)h(H_{ij})
\end{equation}
where $h(z)=h(-z)$ is any even distribution function having finite
second moment. The parameter $p>0$ has a meaning of mean nonzero
elements per row. For $p>1$ the model describes a connected graph
 having locally a tree-like structure
with random connectivity. On a larger scale there exist large loops  
involving
typically of the
order of $\ln{N}$ sites. In the limit $N\to \infty$ the influence of
loops is negligible and the model belongs to the same universality
class as disordered tight-binding model on the infinite tree
(Bethe lattice). The latter model was  thoroughly investigated
by various methods \cite{Abou,FM-bl} and
shown to possess the Anderson transition.

Additional interest to structure of eigenfunctions
on tree-like structures is attributed by a recent paper \cite{AGKL}. 
There, the problem of quasi-particle
lifetime induced by Coulomb interaction in mesoscopic samples
was mapped onto
a disordered tree-like tight binding model in the Fock space,
which is similar to the Bethe lattice and SRM models
and thus undergoes the localization transition.
The fact that in the vicinity of the transition the corresponding
extended eigenfunctions are extremely sparse
 made the authors to conclude that the
 level statistics differs from the WD one for such a regime
because "eigenfunctions do not talk to each other" \cite{AGKL}.
Such a conclusion would be, however, at variance with the results of
explicit calculation of level-level correlation function
performed in the framework of SRM model\cite{sp1}, where it was
shown that
$R_2(\omega)$ is given  by the WD form everywhere in the
region of delocalized states up to the transition point $E=E_c$.

It is necessary to mention that SRM model has some considerable
advantage when compared with more conventional Bethe lattice model.
Namely, quantities like IPR are not
unambiguously defined on the Bethe lattice
in the phase of extended states. Indeed, they require the
consideration of a large but finite lattice to be well defined and their
limiting behavior may depend crucially on boundary conditions
imposed\cite{bound}. In contrast, all sites of SRM model are
essentially equivalent and the model is free from boundary problems.
The general expression for IPR was derived in \cite{sp2} and
its critical behavior analyzed in \cite{FM-ema}.

 Actually, the SRM model
can be used to construct an effective mean-field theory of Anderson
localization\cite{FM-ema} valid at $d=\infty$ \cite{MF-94}.
Critical properties of such theories first discovered on the
level of non-linear $\sigma-$models \cite{Efe-bl,Zirn-bl} turn out to
be quite peculiar. In particular, the coefficient
$C(E)$ diverges close to the
transition point like $C(|E-E_c|\ll E_c)\propto
\exp{\left(\mbox{const} |E-E_c|^{-1/2}\right)}$ \cite{FM-ema}
in contrast to expected power-law behavior for conventional
 $d-$dimensional systems.
The origin of such a critical dependence was explained
in \cite{FM-ema,MF-94} and stems from the fact that $C(E)$ is
determined essentially by the  "correlation
 volume" $V(\xi)$ (i.e. number of sites at a
distance smaller than correlation length $\xi$), which is
exponentially large,
 $V(\xi)\propto\exp(\mbox{const}\,\xi)$, for tree-like structures,  
whereas
$V(\xi)\propto\xi^{d}$ for a $d-$dimensional lattice. Having this
difference in mind, one can translate all the results obtained
in the framework of $d=\infty$ models to their finite-dimensional
counterparts \cite{MF-94}.

To calculate the overlap function defined in Eq.(\ref{1})
we  follow \cite{MB} and use the identity relating
$\alpha(r,E)$ and $\sigma(r,E,\omega)$ to
 advanced and retarded Green functions $G^{R,A}(
r,E)=\sum_{i=1}^N \frac{\left|\Psi_i(r)\right|^2}{E\pm
i\eta-E_i};\quad \eta\to 0^+$:
\begin{eqnarray}
&& 2\pi^2\left[\Delta^{-1}\alpha(r,E)\delta(\omega)+\Delta^{-2}
\tilde{R}_2(\omega)\sigma(r,E,\omega)\right] \label{4}\\
&&= \mbox{Re}\left[\langle
G^R(r,E)G^A(r,E+\omega)  -
 G^R(r,E)G^R(r,E+\omega)\rangle \right] \nonumber
\end{eqnarray}
where $\tilde{R}_2(\omega)$ is non-singular part of the level-level
correlation function:
$R_2(\omega)=\tilde{R}_2(\omega)+\delta(\omega/\Delta)$.
 Let us consider for definiteness the ensemble of real symmetric
SRM, corresponding to systems with unbroken time-reversal invariance. 
For any site index $r=1,...,N$ we introduce one eight-component
supervector $\Phi^{\dagger}=\left(\Phi^{\dagger}_R,
\Phi^{\dagger}_A\right)$  consisting of two
four-component supervectors
$\Phi_{\sigma}^{\dagger}=\left(\phi_{\sigma,b1},\phi_{\sigma,b2},
\phi_{\sigma,f}^*,-\phi_{\sigma,f}\right)$, where indices $\sigma=R,A$
and $b,f$ are used to label advanced-retarded and boson-fermion
subspaces, respectively.
The ensemble-averaged products $\langle G^{\sigma}G^{\sigma'}\rangle$
for RSM model in the limit $N\gg 1$
can be extracted from the paper \cite{sp1} and the
Appendix D of the paper\cite{FM-ema} and is given by:
\begin{eqnarray}
&& \left\langle
G^{\sigma}(r,E)G^{\sigma'}(r,E+\omega) \right\rangle=
\left(1-\frac{4}{3}\delta_{\sigma,\sigma'}\right) \nonumber
\\ \nonumber
&& \times
\int DQ\left\langle \phi_{\sigma,b1}\phi_{\sigma,b1}\phi_{\sigma',b1}
\phi_{\sigma',b1}\right\rangle_{g_T}
\exp{\left(\frac{i\pi\rho\omega N}{4}
\mbox{Str } Q \Lambda\right)}\ ;\\
&&\langle \ldots\rangle_{g_T} = \int
 d\Phi (\ldots) \exp{\left[\frac{i}{2}E\Phi^{\dagger}L\Phi+
pg_T(\Phi)\right]}.\label{5}
\end{eqnarray}
 The function $g_T(\Phi)\equiv g_0(\Phi^{\dagger}
T^{\dagger}T\Phi;\Phi^{\dagger}L\Phi)$ satisfies the integral equation:
\begin{equation}\label{6}
g_T(\Psi)=\left\langle \left[h_F\left(\Phi^{\dagger}L\Psi\right)-1\right]
\right\rangle_{g_T}\ ,
\end{equation}
where $h_F(t)=\int dz e^{-itz}h(z)$ is the Fourier-transform of the
distribution of nonzero elements of the SRM. The
$8\times 8$ supermatrices $T$ satisfy the condition $T^{\dagger}LT=L$
 where $L=\mbox{diag}(1,1,1,1,-1,-1,1,1)$
and belong to a graded coset space whose explicit parametrisation
can be found in \cite{Efrev,VWZ}. The supermatrices $Q$ are expressed
in terms of $T$ as $Q=T^{-1}LT$.  At last, the matrix $\Lambda=
\mbox{diag}(1,1,1,1,-1,-1,-1,-1)$, and the density of states $\rho$
is expressed in terms of the solution of the equation Eq.(\ref{6})
as $\rho(E)=-2g_{0x}/(\pi B_2)$, where $B_2=\int dz h(z) z^2$
and $g_{0x}=\partial g_{0}(x,y)/\partial x|_{x,y=0};\quad
x=\Phi^{\dagger} \Phi,\quad y=\Phi^{\dagger}L \Phi$.

When deriving  Eq.(\ref{5}), evaluation of a functional
integral by the saddle-point method has been employed, see details in
\cite{sp1,FM-ema}. An accurate consideration shows that such a
procedure
is legitimate as long as: i) the matrix size $N$ (playing in our model
the role of the volume) is large enough (much larger than the
coefficient $C(E)$
determining the size dependence of IPR, see above); and
ii) the energy difference $\omega$ is small enough (much smaller than
$C^{-1}(E)$). Though $C(E)$ is exponentially large near the transition
point, it depends on the energy $E$ only, so that when we keep $E$
fixed and increase the system size $N$, the number of levels in the
interval $C^{-1}(E)$ gets arbitrarily large, since the level spacing
scales as $1/N$.

Expanding both sides of Eq.(\ref{6}) over $\Psi$
one can express $\left\langle \phi_{\sigma,b1}
\phi_{\sigma,b1}\phi_{\sigma',b1}
\phi_{\sigma',b1}\right\rangle_{g_T}$ in terms of the matrix $Q$ as
\begin{eqnarray}\nonumber
&& \left\langle \phi_{\sigma,b1}\phi_{\sigma,b1}\phi_{\sigma,b1}
\phi_{\sigma,b1}\right\rangle_{g_T}\\ \nonumber
&&= \frac{4!}{B_4}\left[\frac{1}{2}
g_{0,xx}Q_{b_1b_1}^{\sigma\sigma}Q_{b_1b_1}^{\sigma\sigma}+
 g_{0,xy}Q_{b_1b_1}^{\sigma\sigma}+g_{0,yy}\right]\ ;\\ \nonumber
&& \left\langle \phi_{R,b1}\phi_{R,b1}\phi_{A,b1}
\phi_{A,b1}\right\rangle_{g_T}\\ \nonumber
&& = -\frac{4}{B_4}\left[
g_{0,xx}\left(Q_{b_1b_1}^{RR}Q_{b_1b_1}^{AA}+
2Q_{b_1b_1}^{RA}Q_{b_1b_1}^{AR}\right)+\right. \\
&& \left.
g_{0,xy}\left(Q_{b_1b_1}^{RR}Q_{b_1b_1}^{AA}\right)+g_{0,yy}\right]\ ,
\label{7}
\end{eqnarray}
where $g_{0,xx}=\partial^2 g_0/\partial x^2|_{x,y=0}$;
$g_{0,yy}=\partial^2 g_0/\partial y^2|_{x,y=0}$;
$g_{0,xy}=\partial^2 g_0/\partial x\partial y|_{x,y=0}$,
and $B_4=\int dz h(z) z^4$.
This  allows us to represent right-hand side of Eq.(\ref{4})
in the following form:
\begin{eqnarray} \nonumber
&& 2\pi^2\left[\Delta^{-1}\alpha(r,E)\delta(\omega)+\Delta^{-2}
\tilde{R}(\omega)\beta(r,E,\omega)\right]\\ \nonumber
&&= -\frac{4}{B_4}g_{0xx} \mbox{Re}
\left\langle\left(Q_{b_1b_1}^{RR}Q_{b_1b_1}^{AA}+
2Q_{b_1b_1}^{RA}Q_{b_1b_1}^{AR} \right.\right. \\
&& \left.\left.-\frac{1}{2}\left[
Q_{b_1b_1}^{RR}Q_{b_1b_1}^{RR}+Q_{b_1b_1}^{AA}Q_{b_1b_1}^{AA}\right]\right)
\right\rangle_Q\ ,
\label{8}
\end{eqnarray}
where
$$
\langle ... \rangle_Q=\int dQ (...)
\exp{\left(\frac{i\pi\rho\omega N}{4}
\mbox{Str } Q \Lambda\right)}
$$
The integrals over $Q-$matrices are  the standard ones \cite{Efrev}, 
yielding:
\begin{eqnarray}\nonumber
&& \mbox{Re}\left\langle Q_{b_1b_1}^{RR}Q_{b_1b_1}^{AA}\right\rangle_Q=
1-2R_2^{(0)}(\omega/\Delta)\\ \nonumber
&& \left\langle Q_{b_1b_1}^{RA}Q_{b_1b_1}^{AR}\right\rangle_Q=
-\frac{2i\Delta}{\pi(\omega+i0)};\\
&& \left\langle Q_{b_1b_1}^{RR}Q_{b_1b_1}^{RR}\right\rangle_Q=
\left\langle Q_{b_1b_1}^{AA}Q_{b_1b_1}^{AA}\right\rangle_Q=1\ ,
 \label{9}
\end{eqnarray}
where $R_2^{(0)}(\omega/\Delta)$ is the level correlation function in
the Gaussian Orthogonal Ensemble.
Substituting this in Eq.(\ref{8}), we finally find
\begin{equation}\label{main}
\sigma(r,E,\omega)=
\frac{1}{3}\alpha(r,E)=\frac{1}{N^2}\frac{4g_{0,xx}}
{\pi^2\rho^2 B_4}
\end{equation}

The coefficient $1/3$ in Eq.(\ref{main}) corresponds to the case of
unbroken time reversal symmetry (orthogonal ensemble). For the unitary
ensemble (broken time reversal symmetry) the same consideration yields
the coefficient $1/2$ instead, so that the general relation reads
\begin{equation}
\sigma(r,E,\omega)=\frac{\beta}{\beta+2}\alpha(r,E)\ ,
\label{9a}
\end{equation}
where $\beta$ is the conventional symmetry parameter equal to $\beta=1\
(2)$ for the orthogonal (resp. unitary) ensembles.
This relation between the overlap of two different eigenfunctions
$\sigma(r,E,\omega)$
and self-overlap $\alpha(r,E)$ constitutes the
main result of the  present publication.
It is valid {\it everywhere} in the phase of extended eigenstates, up to
the mobility edge $E=E_c$, provided the number of sites (the system  
volume)
exceeds the correlation volume. In particular, it is valid
in the critical region $|E-E_c|\ll E_c$, where a typical eigenfunction
is very sparse and self-overlap (hence, IPR) grows like
$\exp{\left(\mbox{const} |E-E_c|^{-1/2}\right)}$\cite{FM-ema}.

Eq.(\ref{9a}) implies the following structure of eigenfunctions within
an energy interval
$\delta E=\omega<C^{-1}(E)$.
Each eigenstate can be represented as a product
$\Psi_i(r)=\psi_i(r)\Phi_E(r)$.
The function $\Phi_E(r)$ is an eigenfunction envelope of "bumps and dips"
 which is smooth
on a microscopic scale comparable with lattice constant.  It is the  
same for
all eigenstates around  energy E, reflects underlying gross
(multifractal)
spatial structure and governs the divergence of self-overlap
at critical point. In contrast, $\psi_i(r)$ is Gaussian white-noise
component fluctuating in space on the scale of lattice constant.
It fills in the "smooth" component $\Phi_E(r)$ in an individual way for
each eigenfunction, but is not critical, i.e. is not sensitive to
the vicinity
of the Anderson transition. These Gaussian fluctuations  are
responsible for the factor $\beta/(\beta+2)$ (which is the same as in
the corresponding Gaussian Ensemble) in Eq.(\ref{9a}).

As was already mentioned, this picture is valid in the energy window
$\delta E\sim C^{-1}(E)$ around the energy $E$; the number of
levels in this window being large as $\delta E/\Delta\sim
NC^{-1}(E)\gg 1$ in the thermodynamic limit $N\to\infty$. These states
form a kind of Gaussian Ensemble on a spatially non-uniform
(multifractal for $E\to E_c$) background $\Phi_E(r)$. Since the
eigenfunction correlations are described by the formula (\ref{9a}),
which has exactly the same form as in the Gaussian Ensemble, it is not
surprising that the level statistics has the WD form
everywhere in the extended phase \cite{sp1}.

We believe on physical grounds that the same picture should hold for a
conventional $d$-dimensional conductor. First of all, the general
mechanism of the transition is the same in $d<\infty$ and $d=\infty$
models. Furthermore, the sparsity (multifractality) of eigenstates
near the transition point takes its extreme form for $d=\infty$ models
\cite{MF-94}, so that since the strong correlations (\ref{9a}) take
place at $d=\infty$ it would be very surprising if they do not hold
at finite $d$ as well. Finally, Eq.(\ref{9a}) was proven by an
explicit calculation in the weak localization regime \cite{MB}, where
$\sigma(\bbox{r},E,\omega)={\beta\over\beta+2}\alpha(\bbox{r},E)=
V^{-2}[1+\Pi(\bbox{r},\bbox{r})]$, with $V$ being the system volume and
$\Pi(\bbox{r},\bbox{r})$ the diffusion propagator.

Replacing $C(E)$ by the $d$-dimensional correlation volume
$\sim\xi^d$, we conclude that for $E$ close to $E_c$  eq.(\ref{9a})
should be valid for $\omega<\Delta_\xi$, where $\Delta_\xi\propto
1/\xi^d$ is the level spacing in the correlation volume. For larger
$\omega$, $\sigma(\bbox{r},E,\omega)$ is expected to decrease as
$\omega^{-\eta/d}$ according to the scaling arguments
\cite{Chalk,Huck,Pracz}, so that we find
$\sigma(\bbox{r},E,\omega)/\alpha(\bbox{r},E)\sim
(\omega/\Delta_\xi)^{-\eta/d}$, up to a numerical coefficient of order
of unity. Again, for any value of the energy $E$ in the delocalized
phase, taking the system size $L$ large enough, $L\gg\xi$, we
have a large number of levels
${\delta E/\Delta}\sim{\Delta_\xi/\Delta} \propto (L/\xi)^d$
in the energy window $\delta E$ where eq.(\ref{9a}) holds, so that
the level correlation will be of the WD form.

Finally, let us consider what happens when we go from the critical
regime ($\xi$ large, but $L\gg\xi$) to the critical point ($\xi\gg
L$). For this purpose, let us keep the system size $L$ fixed and
change the energy toward $E_c$, so that $\xi$ increases. When $\xi$
gets comparable to the system size, $\xi\sim L$, we have
$\Delta_\xi\sim\Delta$. This is the border of applicability of the
above consideration. Correspondingly, we find
\begin{equation}
\label{10}
\sigma(\bbox{r},E,\omega)/\alpha(\bbox{r},E)\sim 1\ ,\qquad \omega<\Delta
 \end{equation}
and $\sigma(\bbox{r},E,\omega)/\alpha(\bbox{r},E)\sim
 (\omega/\Delta)^{-\eta/d}$ for $\omega>\Delta$. When $E$ approaches
 further $E_c$, the correlation length $\xi\gg L$ gets irrelevant, so
 that these results will hold in the critical point ($\xi=\infty$). Of
 course, Eq.(\ref{10}) is not sufficient to ensure the WD statistics
in the critical point, since there is only of order of one level
 within its validity range $\delta E\sim\Delta$. Indeed, the
numerical simulations show that the level
 statistics on the mobility edge is different from the WD one
\cite{AS2,Shk,Eva,Braun,Zhar}.

However, eq.(\ref{10}) allows us to make an important conclusion
concerning the behavior of $R_2(\omega)$ at small $\omega<\Delta$, or,
which is essentially the same, the behavior of the nearest neighbor
spacing distribution $P(s)$, $s=\omega/\Delta$, at $s<1$. For this
purpose, it is enough to consider only two neighboring levels. Let
their energy difference be $\omega_0\sim\Delta$. Let us now perturb
the system by a random potential $V(\bbox{r})$ with
$\langle V(\bbox{r})\rangle=0$,
$\langle
V(\bbox{r})V(\bbox{r'})\rangle=\Gamma\delta(\bbox{r}-\bbox{r'})$.
For the two-level system it reduces to a $2\times 2$ matrix
$\{V_{ij}\},\ i,j=1,2$, with  elements $V_{ij}=\int d^d\bbox{r}
V(\bbox{r})
\Psi_i^*(\bbox{r})\Psi_j(\bbox{r})$. The crucial point is that the
variances of the diagonal and off-diagonal matrix elements are
according to Eq.(\ref{10}) equal
to each other up to a factor of order of unity:
\begin{equation}
\langle V_{11}^2\rangle/\langle|V_{12}^2|\rangle =
\sigma(\bbox{r},E,\omega)/\alpha(\bbox{r},E)\sim 1
\label{11}
\end{equation}
The distance between the perturbed levels is given by
$\omega=[(V_{11}-V_{22}+\omega_0)^2+|V_{12}|^2]^{1/2}$. Choosing the
amplitude of the  potential in such a way that the typical energy
shift $V_{11}\sim\Delta$ and using eq.(\ref{11}) and the standard
symmetry consideration, we immediately conclude that in the critical
point $P(s)\simeq c_\beta s^\beta$ for $s\ll 1$ with a coefficient
$c_\beta$ of order of unity, in agreement with the numerical findings
\cite{Shk,Eva,Zhar}.

We are grateful to Y.Gefen and V.E.Kravtsov for stimulating discussions.
The financial support from SFB 195 (A.D.M) and SFB 237 (Y.V.F.)
der Deutschen Forschungsgemeinschaft as well as from the program
 "Quantum Chaos" (grant No. INTAS-94-2058) (Y.V.F.) is acknowledged
 with thanks.


\vspace{-0.5cm}
\begin{thebibliography}{99}
\bibitem{AS2} B.L.Altshuler, I.K.Zharekeshev, S.A.Kotochigova and
B.I.Shklovskii,  Sov.Phys. JETP {\bf 67}, 625 (1988).
\bibitem{Chalk} J.T.Chalker, G.J.Daniell, Phys.Rev.Lett. {\bf 61}, 593
(1988).
\bibitem{Shk} B.I.Shklovskii, B.Shapiro, B.R.Sears, P.Lambrianides,
and H.B.Shore, Phys.Rev.B {\bf 47}, 11487 (1993).
\bibitem{Huck} B.Huckestein and L.Schweitzer,  Phys.Rev.Lett.
{\bf 72}, 713 (1994); T.Brandes, B.Huckestein, L.Schweitzer,
Ann. Physik {\bf 5}, 633 (1996).
\bibitem{Eva} S.N.Evangelou, Phys. Rev. {\bf B 49}, 16805 (1994).
\bibitem{Braun} D.Braun, G.Montambaux, Phys. Rev. {\bf B 52}, 13903  
(1995).
\bibitem{Zhar} I.Kh.Zarekeshev and B. Kramer,
Phys. Rev.  {\bf B 51}, 17356 (1995).
\bibitem{Krav} V.E.Kravtsov,I.V.Lerner,B.L.Altshuler and A.G.Aronov,
  Phys.Rev.Lett. {\bf 72}, 888 (1994);
A. G. Aronov, V. E. Kravtsov, I. V. Lerner,
 Phys. Rev. Lett. {\bf 74}, 1174 (1995).
\bibitem{AM} A.G.Aronov and A.D.Mirlin, Phys.Rev.B {\bf 51}, 6131 (1995).
\bibitem{CLS} J.T.Chalker,I.V.Lerner and R.S.Smith,  Phys.Rev.Lett.
{\bf 77}, 554 (1996); J.Math.Phys. {\bf 37}, 5061 (1996).
\bibitem{CKL}  J.T.Chalker, V.E.Kravtsov, and I.V.Lerner, JETP Lett.
{\bf 64}, 386 (1996).
\bibitem{Pracz} K.Pracz, M.Janssen, P.Freche, preprint cond-mat/9605012.
\bibitem{Weg} F.Wegner, Z.Phys.B {\bf 36}, 209 (1980).
\bibitem{CdC} C.Castellani, C. di Castro, L.Peliti,
              J.Phys. A: Math.Gen. {\bf 19},  L1099 (1986);
\bibitem{Efrev} K.B.Efetov,  Adv. in Phys. {\bf 32}, 53 (1983).
\bibitem{AShk} B.L.Altshuler, B.I.Shklovskii,  Sov.Phys. JETP
 {\bf 64} (1986), 127.
\bibitem{sp1} A.D.Mirlin and Y.V.Fyodorov,  J.Phys.A: Math.Gen.
{\bf 24}, 2273 (1991).
\bibitem{sp2} Y.V.Fyodorov and A.D.Mirlin, Phys.Rev.Lett. {\bf
67}, 2052 (1991).
\bibitem{Abou} R.Abou-Chacra, P.W.Anderson, and D.J.Thouless,  J.Phys.C.
{\bf 6}, 1734 (1973).
\bibitem{FM-bl} A.D.Mirlin, Y.V.Fyodorov,  Nucl.Phys. {\bf B366}
[FS],  507 (1991).
\bibitem{AGKL} B.Altshuler, Y.Gefen, A.Kamenev, and L.Levitov,
preprint  cond-mat/9609132.
\bibitem{bound} Let us remind the reader that for tree-like structures
a macroscopic fraction (of order of unity) of lattice sites
 belong to the boundary.
\bibitem{FM-ema} Y.V.Fyodorov, A.D.Mirlin, H.-J.Sommers,
                  J.Phys.I France {\bf 2}, 1571  (1992).
\bibitem{MF-94} A.D.Mirlin,
Y.V.Fyodorov,  Phys.Rev.Lett. {\bf 72}, 526 (1994);
 J. Phys.I France {\bf 4}, 655 (1994).
\bibitem{Efe-bl} K.B.Efetov,
                Sov.Phys. JETP {\bf 65}, 360 (1987).
\bibitem{Zirn-bl} M.R.Zirnbauer, Phys.Rev. {\bf B 34}, 6394 (1986);
           Nucl.Phys. {\bf B 265}, 375 (1986).
\bibitem{MB} Ya.M.Blanter, A.D.Mirlin, preprint cond-mat/9604139.
\bibitem{VWZ} J.J.M. Verbaarschot, H.-A.Weidenm\"{u}ller, M. R.
Zirnbauer,
 Phys.Rep. {\bf 129}, 367 (1985).
\end{thebibliography}
\end{document}